\pdfoutput=1

\documentclass[11pt]{article}

\usepackage[]{acl2023}

\usepackage{times}
\usepackage{latexsym}

\usepackage[T1]{fontenc}

\usepackage[utf8]{inputenc}

\usepackage{microtype}

\usepackage{inconsolata}
\usepackage{multicol}
\usepackage{booktabs}
\usepackage{colortbl}
\usepackage[fleqn]{amsmath}
\usepackage{multirow}
\usepackage[normalem]{ulem}
\useunder{\uline}{\ul}{}
\usepackage{hyperref}
\usepackage{graphicx}
\usepackage{subfigure}
\usepackage{fixltx2e}
\usepackage{xcolor}
\usepackage{algorithm}
\usepackage{algorithmic}
\usepackage{setspace}
\usepackage{amsfonts}

\title{ChatRetriever: Adapting Large Language Models for Generalized and Robust Conversational Dense Retrieval}

\author{Kelong Mao$^{1}$, Chenlong Deng$^{1}$, Haonan Chen$^{1}$, Fengran Mo$^{2}$,  \\ 
\textbf{Zheng Liu}$^{3}$, \textbf{Tetsuya Sakai}$^{4}$, \textbf{Zhicheng Dou$^{1}$}\thanks{Corresponding author.}, \\
        $^1$Gaoling School of Artificial Intelligence, Renmin University of China \\ 
        $^2$Université de Montréal, Québec, Canada \\ 
        $^3$Beijing Academy of Artificial Intelligence\\
        $^4$Waseda University, Tokyo, Japan\\
        \texttt{\{mkl,dou\}@ruc.edu.cn} \\
}

\begin{document}
\maketitle

\begin{abstract}
Conversational search requires accurate interpretation of user intent from complex multi-turn contexts. This paper presents ChatRetriever, which inherits the strong generalization capability of large language models to robustly represent complex conversational sessions for dense retrieval. 
To achieve this, we propose a simple and effective dual-learning approach that adapts LLM for retrieval via contrastive learning while enhancing the complex session understanding through masked instruction tuning on high-quality conversational instruction tuning data.
Extensive experiments on five conversational search benchmarks demonstrate that ChatRetriever substantially outperforms existing conversational dense retrievers, achieving state-of-the-art performance on par with LLM-based rewriting approaches. Furthermore, ChatRetriever exhibits superior robustness in handling diverse conversational contexts. 
Our work highlights the potential of adapting LLMs for retrieval with complex inputs like conversational search sessions and proposes an effective approach to advance this research direction.



\end{abstract}

\section{Introduction}

Conversational search is rapidly gaining prominence and reshaping how users interact with search engines to foster a more natural information-seeking experience.
At the heart of a conversational search system lie two key components: retrieval and generation~\cite{microsoft22_neural_CIR_survey, arxiv23_llm4ir_survey}. The retrieval process is tasked with sourcing relevant passages, which the generation component then uses to craft the final response.
Conversational retrieval plays a crucial role in ensuring the accuracy and reliability of the system responses by providing relevant passages~\cite{emnlp23_generative_search_engine_eval}.

\begin{figure}[!t]
	\centering
	\includegraphics[width=0.49\textwidth]{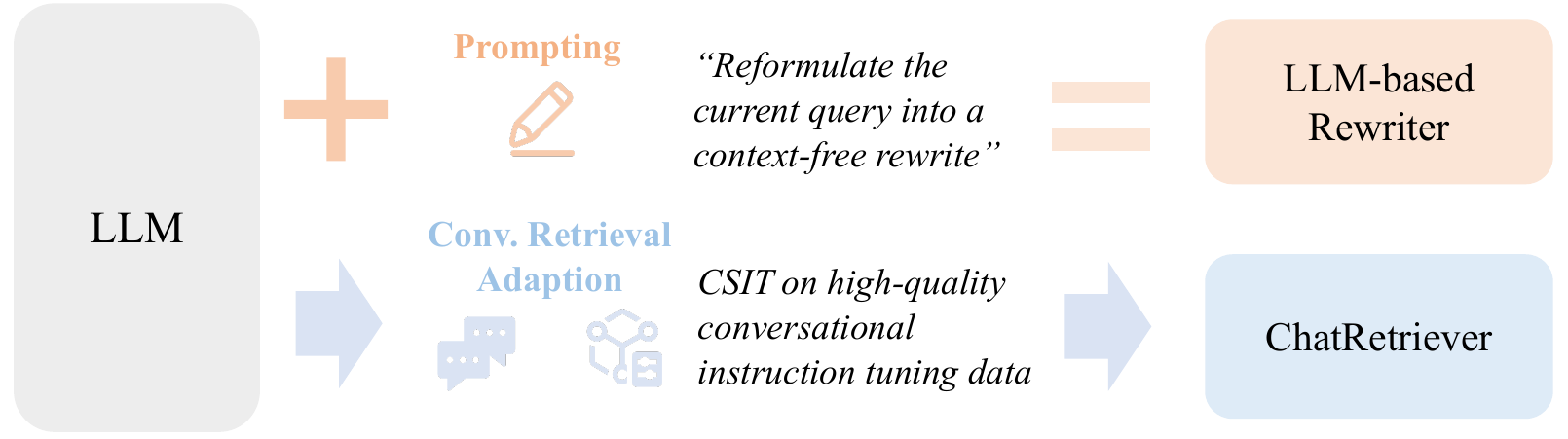}
 
	\caption{Illustration of adapting LLM for query rewriting and conversational dense retrieval.}
	\vspace{-2ex}
	\label{fig:overview}
\end{figure}

Compared to traditional ad-hoc web search, conversational retrieval requires an accurate understanding of the user's real search intent within longer, noisier, and more complex conversational contexts. 
A ``shortcut'' approach is to transform the conversational session into a standalone query rewrite, enabling the usage of ad-hoc retrievers for conversational retrieval. However, the additionally introduced rewriting process is hard to directly optimize towards better retrieval, and it also introduces extra search latency from the rewriting step~\cite{sigir21_ConvDR}.
In contrast, the end-to-end conversational dense retrieval appears to be more promising, as it directly encodes the original conversational search session and passages into dense representations without additional input processing and can enjoy the efficiency benefit from advanced approximate nearest neighbor search algorithms (e.g. Faiss~\cite{bigdata21_faiss}).

Nonetheless, the effectiveness of existing conversational dense retrievers largely trails behind state-of-the-art conversational query rewriting approaches, which leverage large language models (LLMs). Owing to their strong text understanding and generation capabilities, LLM-based rewriters~\cite{emnlp23_llm4cs, emnlp23_EDQR} have demonstrated exceptional effectiveness, even outperforming human rewrites. Given that LLMs are inherently generative models, they can naturally serve as a high-quality conversational rewriter just through prompting (Figure~\ref{fig:overview}). The question that remains is: \textit{whether the potent capabilities of LLMs can be harnessed to substantially enhance the performance of conversational dense retrievers}.

Several studies have explored tuning LLMs for dense retrieval but with a primary focus on ad-hoc search~\cite{acl23_findings_tart, acl23_findings_one_embedder_any_task, arxiv23_repllama, arxiv24_e5_mistrial, arxiv24_Grit_generative_representational_instruction_tuning}. While in conversational search, the multi-turn sessions exhibit greater diversity, complex expressions, and longer-tail intents compared to single-turn ad-hoc queries, posing severe challenges to the session representation learning. Additionally, these approaches often rely on manually designed and fixed instruction templates, which can considerably limit their ability to generalize and handle intricate conversational scenarios.

In this work, we propose adapting LLM itself to serve as a powerful conversational dense retriever.
To achieve this, we select high-quality conversational instruction tuning data~\cite{emnlp23_ultrachat} as our training data and propose a simple dual-learning approach
called \textit{Contrastive Session-Masked Instruction Tuning (CSIT)} for the model training.
Specifically, we adopt the classical contrastive
ranking loss function~\cite{tmlr22_contriever} to fine-tune LLM from a generative model to a retrieval
(or representational) model on the multi-turn instruction (i.e., session)-response pairs, using the special tokens at the end of the input text to represent the entire text.
Meanwhile, we mix the basic contrastive learning with a session-masked instruction tuning objective, where we mask all tokens except the special tokens of the session when computing the language modeling loss of the response tokens. 
The incorporation of this generative instruction tuning loss forces a strong enhancement in the learning of the complex session representation since the response tokens have to be generated solely based on the special tokens representing the session. Furthermore, it also helps retain the strong generalization capability of LLM for retrieval.

Our resulting model, which we call \textbf{ChatRetriever}, can inherit the strong generalization capability of LLM to robustly represent complex conversational sessions for dense retrieval.
 We conducted extensive experiments across five conversational search benchmarks, where ChatRetriever substantially outperforms existing conversational dense retrievers. Notably, it achieves absolute NDCG@3 improvements of 6.8\% and 12.2\% on CAsT-20 and CAsT-21, respectively, matching the performance of the leading LLM-based conversational query rewriting methods. 
Beyond standard evaluations using fixed conversational trajectories, we also developed two robustness evaluation methods to assess the resilience of conversational retrieval approaches by altering the historical context. ChatRetriever demonstrates markedly more stable performance in our robustness test, showcasing its superior robustness in comparison to baselines when faced with varied contexts.

Our contributions can be summarized as:

(1) We introduce ChatRetriever, the first LLM-adapted conversational dense retriever, which substantially outperforms existing conversational dense retrievers and achieves performance comparable to LLM-based rewriting approaches.

(2) We propose Contrastive Session-Masked Instruction Tuning for such a retrieval-oriented adaption for LLM, which can help achieve better complex session representation and generalization.

(3) We design two robustness evaluation methods for conversational retrieval by systematically varying the conversation contexts. 
Results highlight ChatRetriever's superior generalization capability in handling diverse conversational search scenarios.

\section{Related Work}

\noindent \textbf{Conversational search} has seen the development of two primary approaches: conversational query rewriting (CQR) and conversational dense retrieval (CDR). The former approach transforms the conversational search problem into a traditional ad-hoc search problem by reformulating the conversational context into a standalone query. Techniques in this area range from selecting useful tokens from the context~\cite{sigir20_QuReTeC, tois21_HQE} to training generative rewriters based on session-rewrite pairs~\cite{sigir20_autorewriter, emnlp22_conqrr, acl23_findings_edircs, acl23_ConvGQR}.
Inspired by the strong language generation capability of LLMs, some studies~\cite{emnlp23_llm4cs, emnlp23_EDQR, arxiv24_RetPo} propose to leverage LLMs as query rewriters and achieve amazing performance.
Conversational dense retrieval (CDR), on the other hand, directly encodes the entire conversational session for end-to-end dense retrieval~\cite{sigir21_ConvDR}. Efforts in this direction have focused on improving session representation through various perspectives such as context denoising~\cite{sigir22_COTED, kdd23_learning_to_related_to_previous_turns, www2023_lecore}, data augmentation using other corpus and LLMs~\cite{emnlp21_CQE, emnlp22_ConvTrans, icml22_dialog_inpainting, emnlp23_instructor, arxiv24_convaug, www24_workshop_ConvSDG}, and hard negative mining~\cite{emnlp22_saving_shortcut_cdr, arxiv24_haconvdr}.  \\

\begin{figure*}[!t]
	\centering
	\includegraphics[width=1.0\textwidth]{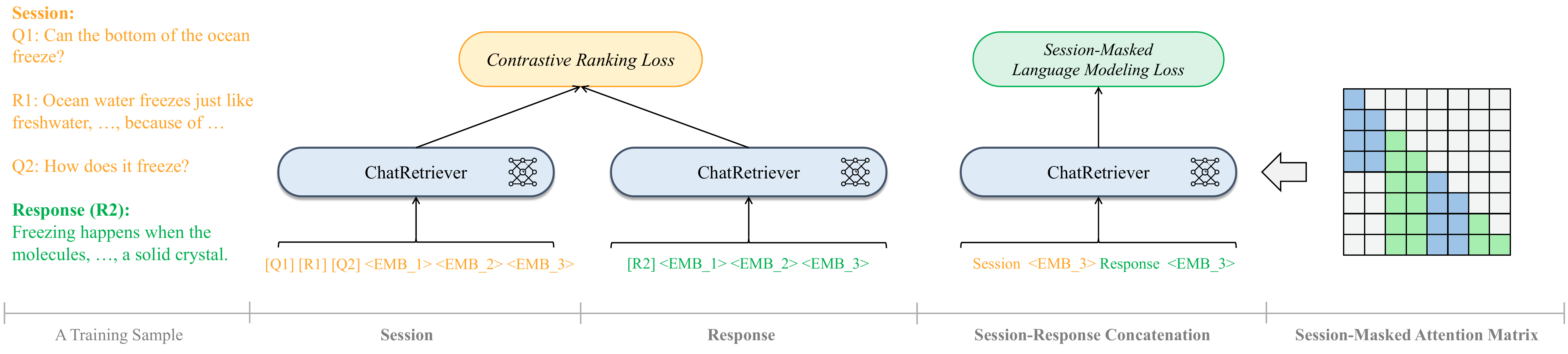}
 
	\caption{Overview of CSIT. We fine-tune LLM to be ChatRetriever using dual learning objectives. We use the last special token (i.e., <EMB\_3>) to represent the input text, which can be \textcolor{orange}{session} or \textcolor[rgb]{0.14, 0.75, 0.46}{response}. In the session-masked attention matrix,  the \textcolor[rgb]{0.62,0.76,0.90}{blue} squares denote the session or the response tokens while the \textcolor[rgb]{0.65, 0.93, 0.69}{green} squares denote their special tokens.}
	\vspace{-2ex}
	\label{fig:model}
\end{figure*}

\noindent \textbf{LLM-based and instruction-aware retrieval.}
Existing research has demonstrated that similar to the scaling laws~\cite{arxiv20_scaling_law_llm} observed in LLMs, increasing the scale of models, data, and computing resources can also enhance the performance of retrieval models~\cite{emnlp22_gtr}. To incorporate the ability to follow instructions into retrievers, some studies~\cite{acl23_findings_one_embedder_any_task, acl23_findings_tart} propose the creation of fixed instruction templates for various retrieval tasks, and use these instruction-enhanced datasets to train the retrievers. Moreover, there have been efforts to adapt LLMs for retrieval purposes by training on improved search data~\cite{arxiv23_repllama, arxiv24_e5_mistrial} or developing new search-oriented training objectives~\cite{arxiv23_llara}. However, these approaches often rely on manually designed and fixed instruction templates, which can limit the generalization capabilities of the retrievers across diverse instructions. Additionally, they are typically designed for single-turn ad-hoc search, lacking the capability to comprehend long and complex search sessions. In contrast to LLMs, which can smoothly understand a wide range of complex user inputs, existing LLM-based retrievers still exhibit a large gap in their generalization capabilities, particularly in the context of conversational search.
\section{Methodology}



We describe our simple and effective dual-learning approach, \textit{Contrastive Session-Masked Instruction Tuning (CSIT)}, which is designed to adapt LLM to a generalized and robust conversational dense retriever. An overview is shown in Figure~\ref{fig:model}.
\\

\noindent \textbf{Contrastive instruction tuning.}
Recent works
have demonstrated the effectiveness of simply using the contrastive ranking loss to adapt LLM to a retriever~\cite{acl23_findings_tart, acl23_findings_one_embedder_any_task,arxiv23_repllama,arxiv24_e5_mistrial, arxiv24_Grit_generative_representational_instruction_tuning}.  However, their generalization capability can be limited as they overfit the
narrow distribution of ad-hoc queries and fixed instruction templates they were trained on. 
We fine-tune LLM on diverse conversational instruction tuning data for more general conversational retrieval adaption.
Specifically, given a training sample $\{(x,y^{+})\}$ from conversational instruction tuning dataset, where $x$ comprises all historical turns and the current instruction (we call $x$ a \textit{session}) and $y$ is the response, we fine-tune LLM with the contrastive ranking loss:
\begin{eqnarray}
    \mathcal{L}_{\text{C}} = -\text{log} \frac{\phi(x, y^{+})}{\phi(x, y^{+}) + \sum_{y^{-} \in D^{-}} \phi(x, y^{-})},
    \label{eq:ranking_loss}
\end{eqnarray}
where $\phi(x, y) = \text{exp}((E(x) \cdot E(y)) / \tau)$, $E(\cdot)$ is the shared text encoder of the retriever. $D^{-}$ is a negative response collection for $x$. 
$\tau$ is a hyperparameter temperature.

To encode text with LLM, we append $t$ special tokens ([EMB$_1$], ..., [EMB$_t$]) to the end of the input text and utilize the representation of the last token ([EMB$_t$]) as the comprehensive representation of the entire text. This approach is analogous to the text-level chain-of-thought (CoT)~\cite{NIPS22_chain_of_thought} for LLMs. We hypothesize that these $t$ consecutive special tokens act as a representational chain-of-thought, expanding and guiding the learning space to achieve a more effective representation. \\



 

\noindent \textbf{Session-masked instruction tuning.}
To enhance the generalized encoding of complex search sessions, we integrate a session-masked instruction tuning objective with the fundamental contrastive learning.
Given a training sample $(x, y^{+})$, we concatenate the instruction and the response to form one input sequence $s$:
\begin{eqnarray}
\begin{split}
    s = [x_1, ..., x_N, [\text{EMB}_1], ..., [\text{EMB}_t], y_1^{+}, \\
    ..., y_M^{+}, [\text{EMB}_1], ..., [\text{EMB}_t]],
\end{split}
\end{eqnarray}
where $x_i$ and $y_i^{+}$ represent the $i$-th token of the session and the response, respectively. $N$ and $M$ denote the total number of tokens in the session and the response, respectively.
We then input this sequence into the LLM to obtain the token representations. Specifically, the representations for the $(N+t)$ \textbf{session} tokens are obtained through a standard auto-regressive process. However, for the subsequent $(M+t)$ \textbf{response} token representations, we mask the $N$ session token representations and allow only the attention of $t$ special session tokens and their preceding response tokens.
We achieve it by applying a customized attention mask matrix illustrated on the right side of Figure~\ref{fig:overview}.
Correspondingly, the loss function of the session-masked instruction tuning is defined as:
\begin{eqnarray}
    \mathcal{L}_{\text{S}} = -\frac{1}{M}\sum_{i=1}^{M}\text{log}p(y_i^{+}|y_1^{+}, ..., y_{i-1}^{+}, \mathbf{x}_{1:t}),
\end{eqnarray}
where $\mathbf{x}_{1:t}$ are the representations of the $t$ session special tokens, which have been contextualized by the $N$ session tokens.

By masking the session text and forcing correct generation for the response tokens, 
we build a closer connection between the session representation and the response token representations.
The model has to perform a more nuanced understanding of the complex session and accurately encode them into the $t$ session special tokens. 

We combine the contrastive instruction tuning and the session-masked instruction tuning to form the final training objective of ChatRetriever:
\begin{eqnarray}
    \mathcal{L} = \mathcal{L}_{\text{C}} + \alpha \mathcal{L}_{\text{S}},
\end{eqnarray}
where $\alpha$ is a hyperparameter to balance the two losses. \\

\noindent \textbf{Discussion.}
Our dual-learning approach CSIT takes inspiration from several notable works in LLM-based retrieval and input compression such as RepLLaMA~\cite{arxiv23_repllama}, E5$_{\text{mistral-7b}}$~\cite{arxiv24_e5_mistrial}, GRIT~\cite{arxiv24_Grit_generative_representational_instruction_tuning}, Gisting~\cite{neurips23_gisting}, and AutoCompressor~\cite{emnlp23_autocompressor}. However, CSIT distinguishes from them in the following key aspects:
(1) RepLLaMA and  E5$_{\text{mistral-7b}}$ primarily focus on contrastive learning using (synthetic) ad-hoc search data with pre-defined instruction templates, which is hard to generalize to complex conversational search scenarios.
(2) GRIT aims to build a unified model for both retrieval and generation, incorporating vanilla instruction tuning and using different training data for its contrastive learning and instruction tuning.
(3) The mechanism of our session-masked instruction tuning shares similarities with Gisting and AutoCompressor, but they are for a completely different target: improving long-context language modeling, not retrieval.
In contrast, CSIT stands out from these works by specifically addressing the challenges of adapting LLM generalized to complex conversational retrieval. 
\section{Experiments}

\subsection{Setup}

\noindent \textbf{Training data.} 
We fine-tune LLM to be ChatRetriever on high-quality conversational instruction tuning datasets.
We select training samples that are informative, diverse, and exhibit information-seeking intents. Our final training data comprises two sources:
(1) The \textit{Question About the World} subset of {UltraChat}~\cite{emnlp23_ultrachat} and  (2) {MSMARCO}~\cite{nips16_msmarco_data} passage ranking dataset.
Ultrachat is a multi-turn instruction tuning dataset while 
MSMARCO can be deemed as a single-turn search-oriented 
instruction tuning dataset by treating the query as the instruction and the positive passage as the response. We find that incorporating MSMARCO is important to improve the basic (ad-hoc) retrieval performance.
\\



\noindent \textbf{Evaluation data and metrics.}
We conduct evaluations on five public conversational search benchmarks, including {QReCC}~\cite{naacl21_qrecc}, {TopiOCQA}~\cite{tacl22_topiocqa}, {CAsT-19}~\cite{trec_cast19}, {CAsT-20}~\cite{trec_cast20}, and {CAsT-21}~\cite{trec_cast21}.
The retrieval corpus sizes of these five datasets are in the tens of millions. 
Among them, the large-scale QReCC and TopiOCQA have training sets, while the other three CAsT datasets are small datasets that only have test sets. 
We mainly report NDCG@3 to evaluate the retrieval performance, as conversational search is more concerned with the top results~\cite{trec_cast20}.  \\

\begin{table*}[!t]
\centering
\renewcommand\arraystretch{1.12}
\setlength{\belowcaptionskip}{10pt}
\setlength{\tabcolsep}{5.5pt}
\scalebox{0.75}{\begin{tabular}{cccccccc}
\toprule
\textbf{Model}                                & \textbf{Base Model}  & \textbf{\#Model Parameter} & \textbf{QReCC} & \textbf{TopiOCQA} & \textbf{CAsT-19} & \textbf{CAsT-20} & \textbf{CAsT-21} \\ \midrule
\multicolumn{8}{c}{\textit{Conversational Query Rewriting}}                                                                                              \\
T5QR                                 & T5-base~\cite{jmlr_T5}     & 250M           & 31.8  & 22.2     & 41.7    & 29.9    & 33.0    \\
ConvGQR                              & T5-base~\cite{jmlr_T5}     & 250M           & 41.0  & 24.3     & 43.4    & 33.1    & 27.3    \\
LLM4CS (REW)                         & ChatGPT-3.5~\cite{chatgpt3_5} & Unknown        & -     & -        & 43.1    & 35.7    & 40.4    \\
LLM4CS (RAR)                         & ChatGPT-3.5~\cite{chatgpt3_5} & Unknown        & -     & -        & 45.3    & 39.5    & 44.9    \\
LLM4CS                               & ChatGPT-3.5~\cite{chatgpt3_5} & Unknown        & -     & -        & \underline{51.5}    & \textbf{45.5}    & \underline{49.2}    \\ 
\midrule
\multicolumn{8}{c}{\textit{LLM-based Retrieval}}                                                                              \\
LLM Embedder                         & BGE~\cite{arxiv23_bge}         & 110M           &  \underline{50.5}     &  22.4        &  36.6       &  15.3       & 31.2        \\
\textsc{InstrcutOR} & GTR-XL~\cite{emnlp22_gtr}      & 1.5B           &  42.3    &   12.3       &      26.8   &       17.3  &   32.4      \\
RepLLaMA                             & LLaMA-2~\cite{arxiv23_llama2}     & 7B             &  31.8      &      15.0    &    31.6     &  18.3       &  32.7       \\
E5$_{\text{mistral-7b}}$                                   & Mistral~\cite{arxiv23_mistrial}    & 7B             &   32.9    &    16.9      & 31.3        &  15.4        &  32.4        \\
GRIT                                   & Mistral~\cite{arxiv23_mistrial}    & 7B             & 33.5       &  17.3        &  30.9       &    19.3      &    33.6      \\
\midrule
\multicolumn{8}{c}{\textit{Conversational Dense Retrieval}}                                                                                              \\
Conv-ANCE                                 & ANCE~\cite{iclr21_ance}
& 110M           & 45.6     & 20.5        & 34.1    & 27.5    & 34.2       \\
ConvDR                               & ANCE~\cite{iclr21_ance}        & 110M           & 35.7  & 26.4     & 43.9    & 32.4    & 37.4      \\
DialogInpainter                      & T5-Large~\cite{jmlr_T5}    & 770M           & -     & -        & 47.0    & 33.2    & -       \\
LeCoRE                               & SPLADE~\cite{sigir22_SPLADE++}      & 110M           & 48.5  & \underline{31.4}     & 42.2    & 29.0    & 32.3    \\
\hline
ChatRetriever                 & Qwen~\cite{qwen}     & 7B             &  \textbf{52.5$^\dagger$}     & \textbf{40.1$^\dagger$}         &  \textbf{52.1$^\dagger$}       &    \underline{40.0$^\dagger$}    &    \textbf{49.6$^\dagger$}     \\ \bottomrule
\end{tabular}}
\caption{Results of the normal evaluation on five conversational search benchmarks. The base models of CQR methods are their rewriters and the model parameters are also counted as the rewriter's parameters. $\dagger$ denotes significant differences to baselines ($p\textless0.05$).
The best results are bold and the second-best results are underlined.}
\label{table:normal_evaluation}
\vspace{-2ex}
\end{table*}

\noindent \textbf{Baselines.}
We compare ChatRetriever against the following three types of retrieval baselines. The first is CQR baselines, including {T5QR}~\cite{arxiv20_t5rewriter}, {ConvGQR}~\cite{acl23_ConvGQR}, and {LLM4CS}~\cite{emnlp23_llm4cs}.
The original LLM4CS has three prompting methods: REW, RAR, and RTR, and it requires multiple rounds of generation, which is time-consuming. For efficiency consideration, we additionally compare with its two single-generation variants based on RAR and REW;
The second is CDR baselines, including {ConvDR}~\cite{sigir21_ConvDR}, {Conv-ANCE}~\cite{www2023_lecore}, {DialogInpainter}~\cite{icml22_dialog_inpainting},  and {LeCoRE}~\cite{www2023_lecore};
The third is the LLM-based retriever baselines, including 
{\textsc{InstructOR}}~\cite{acl23_findings_one_embedder_any_task}, {LLM Embedder}~\cite{arxiv23_llm_embedder}, {RepLLaMA}~\cite{arxiv23_repllama}, {E5$_{\text{mistral-7b}}$}~\cite{arxiv24_e5_mistrial}, and GRIT~\cite{arxiv24_Grit_generative_representational_instruction_tuning}. More baseline details on in Appendix~\ref{sec:baseline_details}. \\

\noindent \textbf{Implementations.}
We initialize ChatRetriever with Qwen-7B-Chat~\cite{qwen} and train it on eight 40G A100 GPUs using LoRA~\cite{iclr22_lora} with a maximum input sequence length of 1024.
The training process involves 2500 steps with a learning rate of 1e-4, a gradient accumulation of 4 steps, a batch size of 64, and 4 hard negatives per sample. 
For consistency, we adopt the \textit{chatml} input format of Qwen-Chat to form the input of ChatRetriever.
We add three special tokens (i.e., \textit{<|extra\_1|>}, \textit{<|extra\_2|>}, and \textit{<|extra\_3|>}) at the end of the instructions and responses.
For baseline comparisons, we adhere to the implementation settings specified in their original papers.
Code is released at \url{https://github.com/kyriemao/ChatRetriever}.

\begin{table*}[!t]
\centering
\renewcommand\arraystretch{1.12}
\setlength{\belowcaptionskip}{10pt}
\setlength{\tabcolsep}{7.3pt}
\scalebox{0.72}{\begin{tabular}{@{}ccccccccccccc@{}}
\toprule
\multirow{3}{*}{\textbf{Model}} & \multicolumn{6}{c}{\textbf{Partial Response Modification}}                                       & \multicolumn{6}{c}{\textbf{Full Context Modification}}                                           \\ \cmidrule(l){2-7} \cmidrule(l){8-13}
                       & \multicolumn{2}{c}{CAsT-19} & \multicolumn{2}{c}{CAsT-20} & \multicolumn{2}{c}{CAsT-21} & \multicolumn{2}{c}{CAsT-19} & \multicolumn{2}{c}{CAsT-20} & \multicolumn{2}{c}{CAsT-21} \\ \cmidrule(l){2-3} \cmidrule(l){4-5} \cmidrule(l){6-7} \cmidrule(l){8-9} \cmidrule(l){10-11} \cmidrule(l){12-13}     
                       & NDCG@3$\uparrow$          & Diff.$\downarrow$         & NDCG@3$\uparrow$          & Diff.$\downarrow$          & NDCG@3$\uparrow$         & Diff.$\downarrow$         & Mean$\uparrow$         & SD$\downarrow$         & Mean$\uparrow$         & SD$\downarrow$         & Mean$\uparrow$         & SD$\downarrow$         \\ \midrule
LLM4CS                 &   50.4            &  1.1           &    43.8           &  1.7           &  49.4             &    0.2         &   49.7           &  1.5           &   44.0            &  1.1           &     48.4          &    1.4         \\
ConvDR                 &   44.3           &  0.4           &    31.0           &  1.4           &   34.8            &    2.6        &            39.3    &        3.4     &       30.2        & 2.6            &   35.8            &   2.9          \\
LeCoRE                 &   44.5            &  2.3            &  25.4             &    3.6         &  29.9              &    2.4         &  42.0             &  1.9           &    28.3           &  2.2           &      31.0         &   2.3         \\ \hline
ChatRetriever       &   52.2            & 0.1            &     39.5          &  0.5           &   48.9            & 0.7            &  51.5             &  1.6           &   45.8            &   1.7          &       48.8        &   1.8          \\ \bottomrule
\end{tabular}}
\caption{Results of the robust evaluation. \textit{Diff.} represents the absolute difference compared to the results in Table~\ref{table:normal_evaluation} and \textit{SD} represents the standard deviation, where a smaller value means more stable. }
\label{table:robust_evaluation}
\vspace{-3ex}
\end{table*}

\subsection{Normal Evaluation}
The retrieval performance comparisons on the five datasets are reported in Table \ref{table:normal_evaluation}. Our proposed ChatRetriever outperforms all the baseline methods across these datasets. Existing conversational dense retrievers are constrained by limited model capacity and data quality, resulting in suboptimal performance for conversational retrieval tasks.
Prior to ChatRetriever, there was a considerable performance gap between existing conversational dense retrieval methods and the state-of-the-art LLM-based conversational query rewriter (i.e., LLM4CS). Specifically, the absolute gaps between the best existing CDR model and LLM4CS were 1.6\%, 12.2\%, and 11.8\% on the three CAsT datasets, respectively.
However, ChatRetriever can achieve comparable or even superior performance to LLM4CS, highlighting the high potential of end-to-end conversational dense retrieval compared to the two-stage approach of conversational query rewriting methods. If we force LLM4CS to generate a single output (RAR) or only consider query rewriting (REW) for efficiency, the advantages of ChatRetriever become even more pronounced, with over 4\% absolute gains.
We also observe that existing LLM-based retrievers do not perform well on conversational retrieval tasks. This can be attributed to the fact that they are fine-tuned solely on templated instructions, which fails to fully leverage the generalization capabilities of LLMs to handle complex and diverse conversational scenarios.



\subsection{Robustness Evaluation}
Existing evaluations for conversational retrieval are mainly conducted on fixed conversation trajectories.
In this section, we evaluate the robustness of conversational retrievers in different contexts.
Our principle is modifying the context but fixing the current query (i.e., search intents) for each turn so that the original relevance labels can be re-used. 
Specifically, we propose the following two types of context modification: 

(1) \textit{Partial response modification:}
We do not use the provided responses in the evaluation dataset. Instead, for each turn, we input the current query, the context, and the top-3 passages retrieved by the conversational retriever, and prompt LLM to generate the response.
The simulated online nature of generating responses turn-by-turn better matches how conversational retrieval systems are used in practice.
However, a problem with this online evaluation manner is that the query of the next turn in the original dataset may become unreasonable after modifying its last response~\cite{acl22_ditch_the_gold}. 
We propose a simple heuristic method to tackle this problem with LLM.
Specifically, we prompt LLM to judge whether the current query is reasonable given the context.
If not, we replace the current query with its human rewrite to make it stand on its own without needing external context.
Otherwise, we can use the original query. The prompts can be found in Appendix~\ref{sec:prompts_in_partial_response_modification}.

\begin{table*}[!t]
\centering
\renewcommand\arraystretch{1.12}
\setlength{\belowcaptionskip}{10pt}
\setlength{\tabcolsep}{15pt}
\scalebox{0.82}{
\begin{tabular}{ccccccc}
\toprule
\textbf{Base LLM}      & \textbf{Model Parameter} & \textbf{Base/Chat} & \textbf{Training} & \textbf{CAsT-19} & \textbf{CAsT-20} & \textbf{CAsT-21} \\ \midrule
Qwen  & 1.8B       & Chat      & Full      &  38.8       & 33.7        & 45.2        \\
Qwen  & 1.8B       & Chat      & LoRA      &  35.1       &   31.9      &    42.4     \\
Qwen  & 7B         & Base      & LoRA      &   46.9      &   37.7      &  46.5       \\
Qwen  & 7B         & Chat      & LoRA      &  50.5       &  40.0       & 49.6        \\
LLaMA-2   & 7B         & Chat      & LoRA      &  47.3       &  38.4       & 49.1        \\
Mistrial & 7B         & Chat      & LoRA      &  49.5       &  39.2       &  49.6       \\ \bottomrule
\end{tabular}}
\caption{Performance comparisons of ChatRetrievers under different settings with different backbone LLMs.}
\label{table:influence_of_llm}
\vspace{-2ex}
\end{table*}


\begin{table}[!t]
\centering
\scalebox{0.84}{\begin{tabular}{cccc}
\toprule
    Ablation            & CAsT-19 & CAsT-20 & CAsT-21 \\ \midrule
w/o SIT         & 49.5    & 36.8    & 45.8    \\
w/o R-CoT       & 49.9    & 38.5    & 47.5    \\
with Vanilla IT & 51.1    & 39.3    & 48.4    \\
CSIT            & 52.1    & 40.0    & 49.6    \\ \bottomrule
\end{tabular}}
\caption{Results of ablation studies.}
\label{table:ablation_study}
\vspace{-2ex}
\end{table}

(2) \textit{Full context modification:} 
For each turn, we supply the original query and its human-modified version to the LLM, prompting it to generate new contexts (See Appendix~\ref{sec:prompts_in_full_ctx}).
We finally got five different contexts for each turn.

We evaluate conversational retrievers based on different contexts generated by these two modification methods using ChatGPT 3.5.
For the partial response modification setting, we report the retrieval performances and their absolute differences (\textit{Diff.}) compared to the original counterpart results reported in Table~\ref{table:normal_evaluation}.
For the full context modification setting, we report the \textit{Mean} performance of different runs and their \textit{standard deviation (SD)}. 
The robust evaluation results are shown in Table~\ref{table:robust_evaluation}.

For the partial response modification setting, it shows that the performance changes of ChatRetriever are the smallest. By referring to Table~\ref{table:normal_evaluation}, we also observe a general degradation in retrieval performance compared to the original context. This degradation may stem from the retrieved passages being inaccurate, consequently leading to inaccurate responses, and then affecting the retrieval performance of the subsequent turns.

For the full context modification setting, the robustness of ChatRetriever is further highlighted by its small average standard deviation of 1.7, which is lower compared to the 3.0 and 2.1 standard deviations observed for ConvDR and LeCoRE, respectively. These results demonstrate the strong robustness of ChatRetriever to different conversational search contexts.
In contrast, the LLM4CS, which utilizes ChatGPT for query rewriting, shows an even lower standard deviation of 1.3,  demonstrating the superior robustness of ChatGPT for conversational query rewriting.



\subsection{Ablation Studies}
We build four ablations to study the effects of our proposed training approach:
(1) \textit{w/o R-CoT}: removing the representational CoT;
(2) \textit{w/o SIT}: removing the session-masked instruction tuning;
(3) \textit{with Vanilla IT}: replacing the session-masked instruction tuning with vanilla instruction tuning.

Table~\ref{table:ablation_study} shows the ablation results.
We find that either removing the representational CoT or removing or replacing session-masked instruction tuning can lead to performance degradation.
By contrast, the session-masked instruction tuning, which achieves 6.6\% relative performance gains across the three CAsT datasets on average, is shown to be more effective than representational CoT, which achieves 3.4\% relative performance gains on average.
The results suggest that our two techniques have positive effects in helping adapt LLMs for conversational retrieval.
We also studied the influence of the number of special CoT tokens, which can be found in Appendix~\ref{sec:influence_of_number_of_special_tokens}.

\begin{table*}[!t]
\centering
\scalebox{0.84}{\begin{tabular}{ccccccccccc}
\toprule
\multirow{2}{*}{Methods} & \multicolumn{2}{c}{QReCC} & \multicolumn{2}{c}{TopiOCQA} & \multicolumn{2}{c}{CAsT-19} & \multicolumn{2}{c}{CAsT-20} & \multicolumn{2}{c}{CAsT-21} \\ \cmidrule(l){2-3} \cmidrule(l){4-5} \cmidrule(l){6-7} \cmidrule(l){8-9} \cmidrule(l){10-11}    
                         & Original      & New       & Original        & New        & Original       & New        & Original       & New        & Original       & New        \\ \midrule
GRIT                     & 33.5          & 48.3      & 17.3            & 36.0       & 30.9           & 47.1       & 19.3           & 35.7       & 33.6           & 45.3       \\
Conv-ANCE                & 45.6          & 44.8      & 20.5            & 21.6       & 34.1           & 35.0       & 27.5           & 30.5       & 34.2           & 36.0       \\
ConvDR                   & 35.7          & 36.0      & 26.4            & 24.9       & 43.9           & 43.2       & 32.4           & 30.9       & 37.4           & 35.5       \\
LeCoRE                   & 48.5          & 46.1      & 31.4            & 31.0       & 42.2           & 42.9       & 29.0           & 30.1       & 32.3           & 33.4       \\ \hline
ChatRetriever            & \multicolumn{2}{c}{52.5}  & \multicolumn{2}{c}{40.1}     & \multicolumn{2}{c}{52.1}    & 40.0           &            & \multicolumn{2}{c}{49.6}    \\ \bottomrule
\end{tabular}}
\caption{Results of continually fine-tuning baselines on the training data of ChatRetriever. ``Original'' and ``New'' denote the performance before and after fine-tuning, respectively.}
\label{table:continuelly_fine_tuning}
\end{table*}



\begin{table}[!t]
\centering
\scalebox{0.84}{\begin{tabular}{ccccc}
\toprule
\multirow{2}{*}{Data Source} & \multicolumn{2}{c}{CAsT-20} & \multicolumn{2}{c}{CAsT-21} \\ \cmidrule(l){2-3} \cmidrule(l){4-5}
                             & Session      & Rewrite      & Session      & Rewrite      \\ \midrule
Only U               & 39.5         & 43.7         & 46.5         & 50.0         \\
Only M                 & 18.3         & 49.8         & 34.1         & 58.9         \\
Q+M                & 31.5         & 46.9         & 42.4         & 47.9         \\
U+M            & 40.0         & 49.9         & 49.6         & 59.2         \\ \bottomrule
\end{tabular}}
\caption{Comparisons of using different data sources combinations for training. U, M, and Q represent UltraChat, MSMARCO, and QReCC, respectively.}
\label{table:data_sources}
\vspace{-2ex}
\end{table}

\begin{figure*}[!t]
	\centering
 \includegraphics[width=0.97\textwidth]{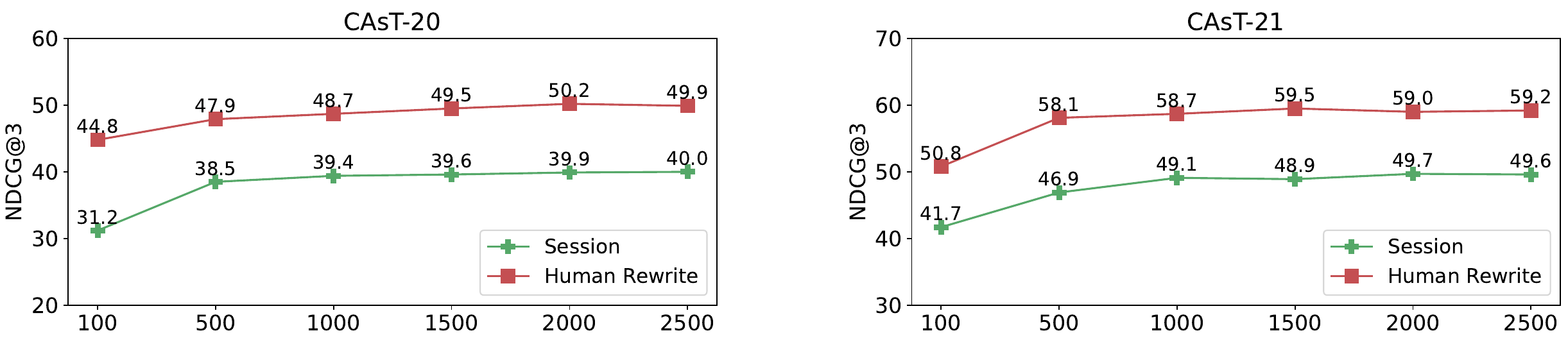}
	\caption{Performance of ChatRetriever at different training steps.}
	\vspace{-2ex}
	\label{fig:data_volume}
\end{figure*}

\subsection{Influence of LLMs}
\label{sec:influence_of_llms}
Table~\ref{table:influence_of_llm} shows the comparisons between different settings about the backbone LLM of ChatRetriever.

(1) \textbf{Base vs. Chat.}
Our results indicate that the Chat model outperforms the Base model, which aligns with our expectations. We hypothesize that the ability to follow instructions well is indicative of strong generalization capabilities, which are crucial for complex conversational search tasks. Therefore, the Chat model, having been fine-tuned for conversational instructions, provides a more appropriate foundation for this task.

(2) \textbf{Different LLMs.}
We find that different LLMs have similar performance under our training recipe. The relatively worst variation based on LLaMA-2 still largely outperforms existing conversational dense retrieval baselines on the more complex CAsT-20 and CAsT-21 datasets, and also outperforms smaller ChatRetrievers.

(3) \textbf{LoRA vs. full parameter tuning.}
Due to constraints in computing resources, our investigation into training modes (i.e., LoRA vs. full parameter tuning) was limited to the 1.8B scale model. Our findings indicate that employing LoRA training yields inferior performance compared to full parameter tuning. However, this may be attributed to the LoRA parameter capacity being insufficient for the 1.8B model.


\subsection{Influence of Training Data}

\noindent \textbf{Fine-tuning on different data sources.}
Table~\ref{table:data_sources} presents the performance of ChatRetriever when trained solely on UltraChat, solely on MSMARCO, and on a combination of QReCC+MSMARCO (i.e., replacing UltraChat with the QReCC's training set). The model performance is evaluated using both session inputs and human rewrite inputs (i.e., converted to ad-hoc search). We find that training exclusively on UltraChat leads to a decline in performance for both input types, with a more pronounced degradation observed for the rewrite input. Conversely, training solely on MSMARCO yields comparable results for the rewrite input but considerably worse performance for the session input. These results suggest that MSMARCO effectively enhances the ad-hoc retrieval capabilities of LLMs, possibly due to its well-curated hard negatives. However, ad-hoc search data from MSMARCO alone is insufficient for transferring the generalization capability of LLMs to the more complex context of conversational search. 
The traditional conversational QA data (i.e., QReCC) is also not highly effective for LLMs in learning a diverse range of complex conversational patterns. To optimize LLM to be a universal conversational retriever, we recommend combining general conversational instruction tuning data (e.g., UltraChat) with ad-hoc search-oriented instruction tuning data (e.g., MSMARCO). 
\\

\noindent \textbf{Continuelly fine-tuning baselines on the same training data of ChatRetriever.}
In Table~\ref{table:normal_evaluation}, we follow the original training settings of the baselines. 
Here, we further fine-tune baselines on the training data of ChatRetriever.
Results are shown in Table~\ref{table:continuelly_fine_tuning} and we find:
(1) GRIT, a unified retrieval and generation model based on LLM, showed substantial performance improvement after fine-tuning on conversational instruction tuning data. Its performance approached that of ChatRetriever without session-masked instruction tuning, although it still lagged behind the final ChatRetriever. 
(2) The performance of Conv-ANCE, ConvDR, and LeCoRE did not show noticeable improvements and even experienced declines in QReCC and TopiOCQA. This may be because that the newly introduced training data disrupted their original in-domain training-test settings, as they were initially trained on the in-domain training sets of QReCC and TopiOCQA.
This also highlights the robust generalization of ChatRetriever, which, when trained only on general conversational instruction tuning data, can effectively adapt to various conversational search test sets. \\


\noindent \textbf{Data volume.}
Figure~\ref{fig:data_volume} shows the performance of ChatRetriever across various training steps.
It is observed that the performance attains a relatively high level at 500 steps and subsequently experiences marginal improvements as the number of training steps increases. The performance stabilizes upon reaching 2500 steps. Furthermore, the trends for inputs with sessions and human rewrites are similar. These findings suggest that, under our framework, adapting LLMs to function effectively as conversational retrievers may require only a small amount of high-quality data.


\section{Conclusion}
In this paper, we introduce ChatRetriever, a large conversational retrieval model adapted from LLM. We propose a novel contrastive session-masked instruction tuning approach for this adaptation and fine-tune LLM on high-quality conversational instruction tuning data. Experimental results on five conversational retrieval datasets demonstrate the superior performance and robustness of ChatRetriever.
Looking ahead, we aim to further explore and expand the generalization capabilities of ChatRetriever in a broader range of complex IR scenarios beyond conversational search, such as legal case retrieval, product search, and other instruction-followed search tasks. We envision ChatRetriever to be as versatile as LLMs, capable of accepting and understanding any conversational inputs and retrieving useful information for those inputs.

\section*{Limitations}

\noindent \textbf{Efficiency.} 
As indicated in Table~\ref{table:normal_evaluation}, ChatRetriever is a 7B model which is much larger than existing CDR models. Our preliminary findings (Section~\ref{sec:influence_of_llms}) suggest that the large model size is a crucial factor for ChatRetriever's exceptional performance. However, this also raises efficiency concerns. With an embedding dimension of 4096, ChatRetriever incurs higher time and storage costs for indexing and retrieval than existing CDR models. Nevertheless, on the one hand, ChatRetriever's enhanced retrieval accuracy potentially reduces the need for extensive passage re-ranking, which could, in real-world applications, offset the initial higher costs by ultimately reducing the total time spent on ranking.
On the other hand, we view ChatRetriever as a promising research direction in leveraging the potent capabilities of LLMs for more complex and potentially universal retrieval tasks. We are exploring the possibility of distilling ChatRetriever into a more efficient, smaller model. \\

\noindent \textbf{Hard Negatives.} Unlike typical search datasets that provide a large retrieval corpus, the conversational instruction tuning dataset we used (i.e., UltraChat) consists of only multi-turn instructions (i.e., sessions) and responses. In this work, we simply chose the CAsT-21 corpus for the hard negative mining of UltraChat (see Appendix~\ref{sec:appendix_hard_neg}). However, as existing studies have shown, hard negatives are crucial for improving retrieval performance~\cite{sigir21_ADORE, emnlp22_SimANS}. Therefore, a better strategy for mining hard negatives tailored to instruction tuning data is desirable. We plan to explore using LLMs to generate hard negatives for instructions similar to \citep{arxiv24_e5_mistrial}. \\

\noindent \textbf{Generalizability.} ChatRetriever substantially outperforms existing CDR models in understanding and retrieving information for complex multi-turn inputs and achieves comparable performance to state-of-the-art LLM-based rewriting, showcasing its strong generalization capability.
However, it has not yet achieved the same level of generalization as LLMs, particularly in following complex retrieval instructions, addressing very detailed information needs, or performing in-context learning across various specific domains. It is worth noting that existing instruction-aware retrievers~\cite{acl23_findings_one_embedder_any_task, arxiv23_llm_embedder, arxiv24_Grit_generative_representational_instruction_tuning} also have limitations in perceiving complex (multi-turn) instructions that largely fall short of the generality of LLMs, as highlighted in this work (Table~\ref{table:normal_evaluation}) and also in recent studies \cite{arxiv24_InstructIR, arxiv24_FollowIR}. As stated in our conclusion, we are committed to further advancing ChatRetriever's generalization capabilities to match those of LLMs. \\





\bibliography{custom}

\appendix

\clearpage
\section*{Appendix}

\section{More Details of Experimental Setup}
\label{sec:baseline_details}

\subsection{Evaluation Datasets}
The basic statistics of these five evaluation datasets are shown in Table~\ref{table:dataset}.
All the datasets except TopiOCQA provide the human rewrite for each turn.
The relevance annotations in the CAsT datasets are made by experts, making them more detailed.

\begin{table}[H]
\centering
\scalebox{0.65}
{\begin{tabular}{@{}cccccc@{}}
\toprule
\textbf{Statistics}   & \textbf{QReCC}  & \textbf{TopiOCQA} & \textbf{CAsT-19}    & \textbf{CAsT-20}    & \textbf{CAsT-21} \\ \midrule
\#Conversation & 2,775  & 205      & 50         & 25         & 26      \\
\#Turns        & 16,451 & 2,514    & 479        & 208        & 239     \\
\#Passages       & 54M    & 25M      & \multicolumn{2}{c}{38M} & 40M     \\ \bottomrule
\end{tabular}}
\caption{Basic statistics of the five evaluation datasets.}
\label{table:dataset}
\end{table}

\subsection{Baselines}
We provide a more detailed introduction to the baselines:

\textbf{T5QR}~\cite{arxiv20_t5rewriter}: a T5-based query rewriting method trained with human rewrites as the supervised signals.

\textbf{ConvGQR}~\cite{acl23_ConvGQR}: A unified framework for query reformulation that integrates rule-based query rewriting with a generative model to expand queries.

\textbf{LLM4CS}~\cite{emnlp23_llm4cs}: A state-of-the-art LLM-based prompting method for conversational query rewriting. LLM4CS has two three prompting methods: REW, RAR, and RTR. REW only generates a rewrite and RAR additionally generates a hypothetical response. While RAR generates a rewrite and response in a two-step manner.
For LLM4CS (REW) and LLM4CS (RAR), we only generate once for efficiency consideration and thus do not need aggregation.

\textbf{Conv-ANCE}~\cite{www2023_lecore}, which uses the classical ranking loss to train the session embeddings based on ANCE~\cite{iclr21_ance}.

\textbf{ConvDR}~\cite{sigir21_ConvDR}, which uses knowledge distillation to learn the session embeddings from rewrites.

\textbf{DialogInpainter}~\cite{icml22_dialog_inpainting}, which is fine-tuned from the T5-large model using  information seeking dialogues generated from large web corpora.

\textbf{LeCoRE}~\cite{www2023_lecore}, which extends SPLADE~\cite{sigir22_SPLADE++}  to be a conversational lexical retriever using multi-level denoising methods.

\textbf{\textsc{InstructOR}}~\cite{acl23_findings_one_embedder_any_task}, a general retriever tailored to various tasks and domains by trained with various task-specific instructions.

\textbf{LLM Embedder}~\cite{arxiv23_llm_embedder}: a unified retrieval model that can support diverse retrieval augmentation needs of LLMs. It is fine-tuned on various tasks and datasets such as MSMARCO, NQ, ToolLLM, QReCC, FLAN, Books3, and Multi-Session Chat.

\textbf{RepLLaMA}~\cite{arxiv23_repllama}, a large ad-hoc retriever fine-tuned from LLaMA-7B on the MSMARCO dataset.

\textbf{E5$_{\text{mistral-7b}}$}~\cite{arxiv24_e5_mistrial}, a large ad-hoc retriever fine-tuned from Mistral-7B on the synthetic dataset generated by ChatGPT and MSMARCO.

\textbf{GRIT}~\cite{arxiv24_Grit_generative_representational_instruction_tuning}, a unified model for retrieval and generation. It is fine-tuned based on Mistral-7B. The retrieval part is fine-tuned on the E5~\cite{arxiv24_e5_mistrial} dataset with task-specific instructions while the generation part is fine-tuned on the Tulu 2~\cite{arxiv23_tulu2} dataset.

\subsection{Hard Negatives}
\label{sec:appendix_hard_neg}
For UltraChat, we first use in-context learning with Qwen-7B-Chat, similar to the approach in \citep{emnlp23_llm4cs}, to generate a query rewrite for each turn. We then obtain hard negatives by randomly sampling from the top-15 to top-30 retrieval results using the LLM Embedder on the CAsT-21 corpus with rewrites. The hard negatives for MSMARCO are consistent with those used in \citep{arxiv23_repllama}.

\begin{figure}[!t]
	\centering
	\includegraphics[width=0.49\textwidth]{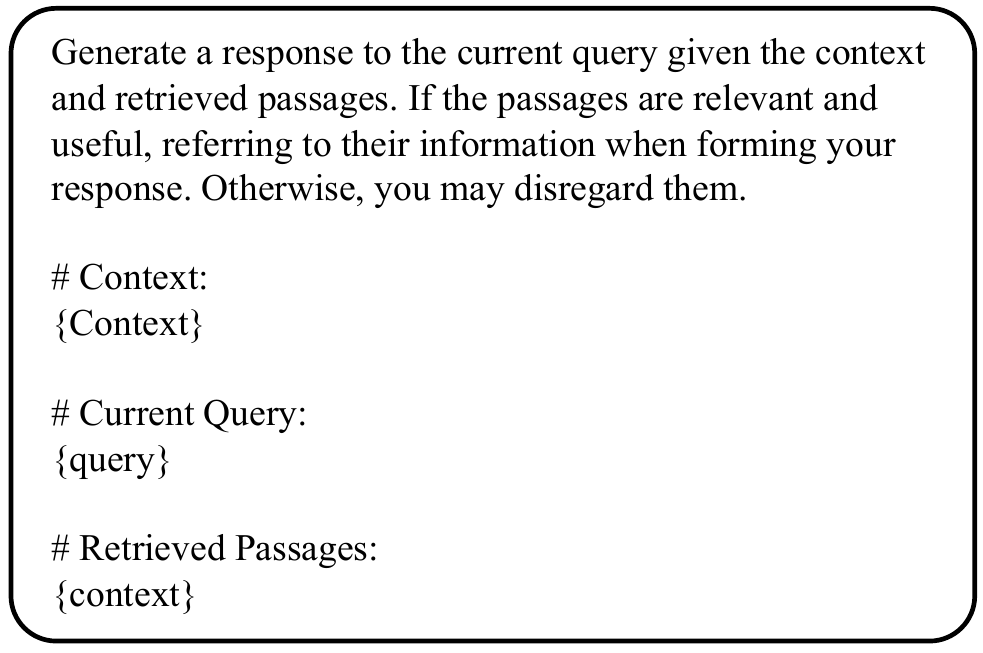}
	\caption{The prompt to generate the response in the experiment of partial response modification.}
	\vspace{-2ex}
	\label{fig:prompt_rag}
\end{figure}

\begin{figure}[!t]
	\centering
	\includegraphics[width=0.49\textwidth]{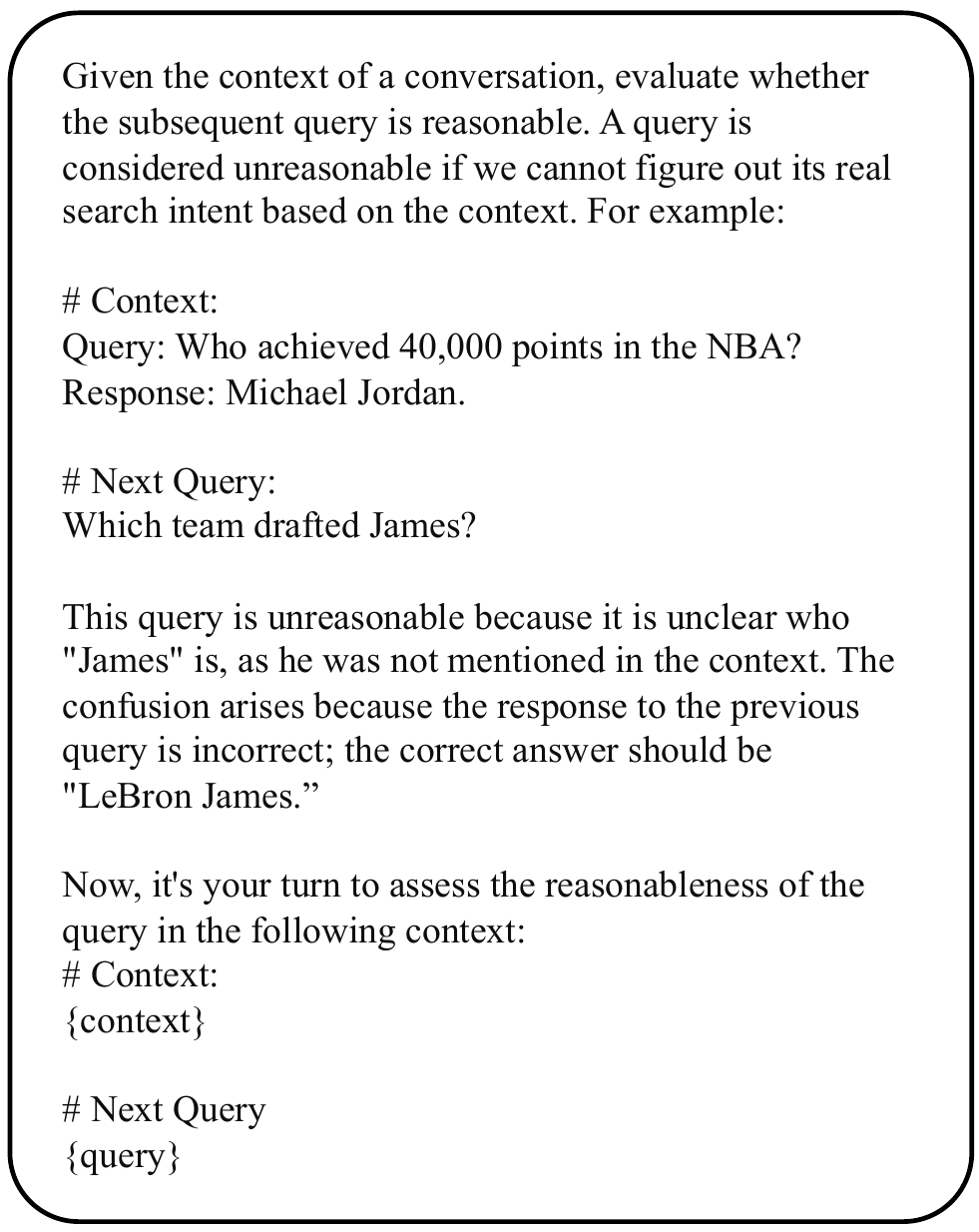}
	\caption{The prompt to judge whether the current query is reasonable in the experiment of partial response modification.}
	\vspace{-2ex}
	\label{fig:prompt_judge}
\end{figure}


\begin{figure*}[!t]
	\centering
	\includegraphics[width=1.0\textwidth]{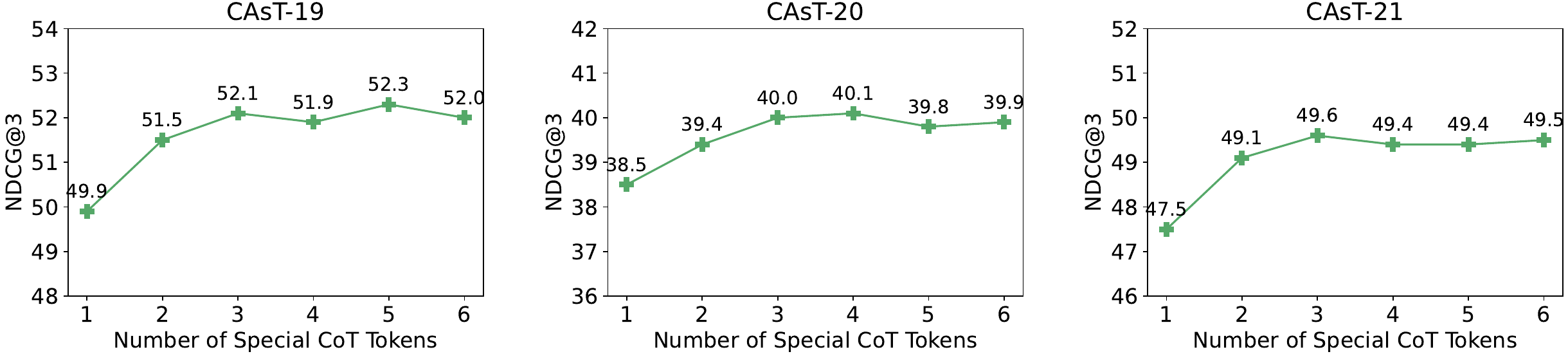}
	\caption{Performance comparisons when using different numbers of special CoT tokens.}
	\vspace{-2ex}
	\label{fig:num_of_tokens}
\end{figure*}

\section{Prompts in Partial Response Modification}
\label{sec:prompts_in_partial_response_modification}
The prompts to generate the response and judge whether the current query is reasonable are shown in Figure~\ref{fig:prompt_rag} and Figure~\ref{fig:prompt_judge}, respectively.

\begin{figure}[H]
	\centering
	\includegraphics[width=0.49\textwidth]{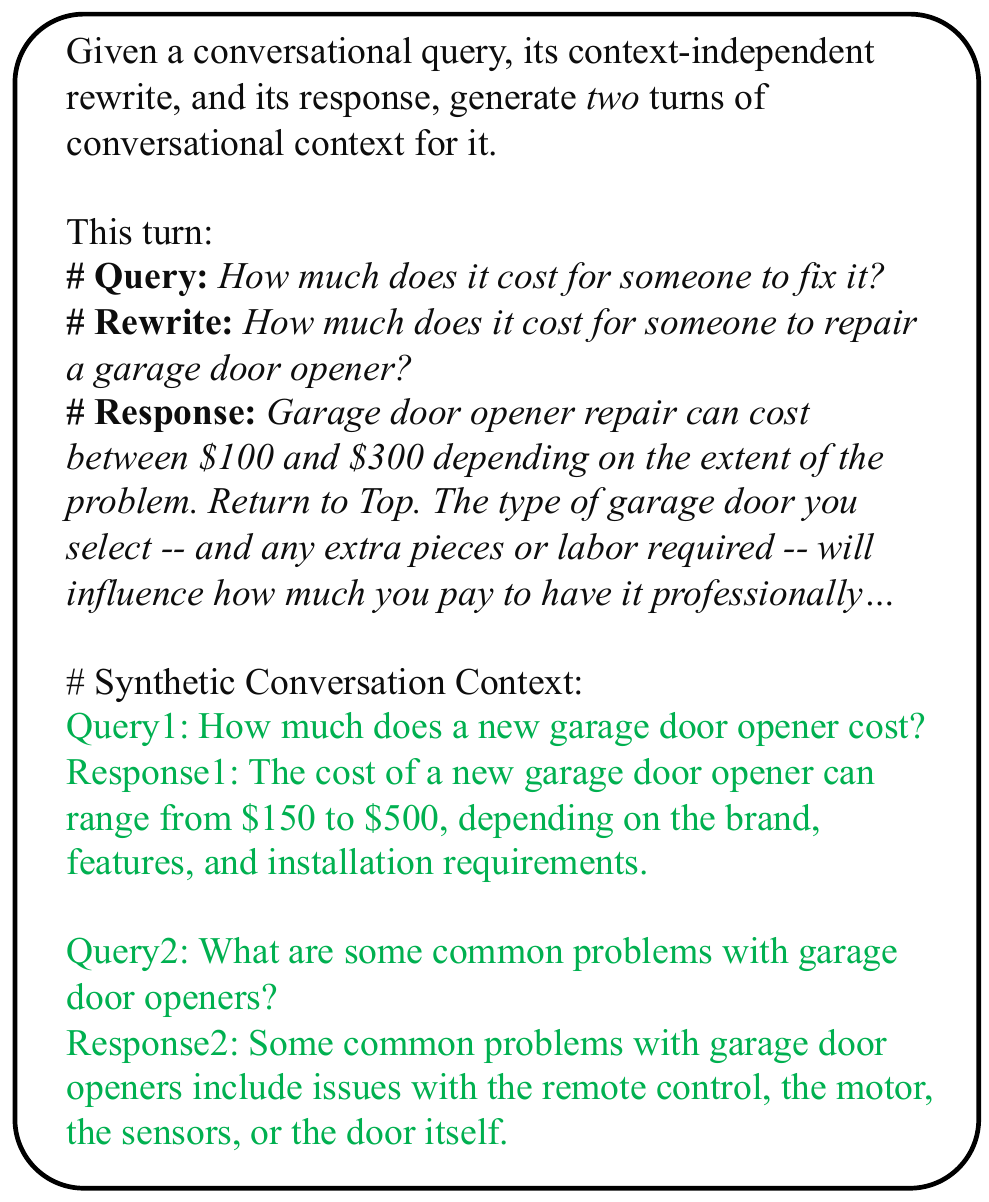}
	\caption{An example prompt to generate synthetic conversation text in the experiment of full context modification. \textit{Italicized contents} are filled into the placeholders of the prompt. The green content is the model output.}
	\vspace{-2ex}
	\label{fig:prompt_full_ctx_modification}
\end{figure}

\section{Prompts in Full Context Modification}
\label{sec:prompts_in_full_ctx}

The prompt to generate synthetic conversation text in the experiment of full context modification is shown in Figure~\ref{fig:prompt_full_ctx_modification}.
The green content is the output of ChatGPT3.5 for the above prompt.

\section{Influence of the Number of Special CoT Tokens}
\label{sec:influence_of_number_of_special_tokens}
In Figure~\ref{fig:num_of_tokens}, we present the performance of ChatRetriever when varying the number of special tokens used for text representation. Our findings suggest that the inclusion of additional special tokens generally enhances retrieval performance. This improvement may be attributed to the fact that a sequence of consecutive special tokens can serve as a form of representational-level CoT, effectively expanding the learning space. However, we observe that performance plateaus when the number of special tokens exceeds three. Consequently, we finally append three special tokens in our implementation.


\section{Settings of Continuelly Fine-tuning Baselines}
Since the training data of ChatRetriever only contains session-response pairs but does not contain human rewrites, we use in-context learning with Qwen-7B-Chat, similar to the approach in \citep{emnlp23_llm4cs}, to generate query rewrite for each turn and use them for the training of ConvDR and LeCoRE.
GRIT and Conv-ANCE are fine-tuned with their original contrastive ranking loss.


\end{document}